\documentclass[sigconf]{acmart}

\usepackage{booktabs}
\usepackage{pbox}
\usepackage{subfigure}
\usepackage[inline]{enumitem}
\usepackage{multirow}
\usepackage{breakurl} 
\usepackage{url}
\renewcommand\footnotetextcopyrightpermission[1]{} 
\usepackage{xspace}
\newcommand{\etal}{{et al.\@\xspace}}

\newcommand{\rec}{{LotRec\xspace}}

\setlist{nosep}
\setdescription{leftmargin=1em}

\begin{document}

\title{\rec: A Recommender for Urban Vacant Lot Conversion}


\author{Md Towhidul A. Chowdhury}
\affiliation{%
 \institution{Rochester Institute of Technology}
 \city{Rochester} 
 \state{NY} 
 \postcode{14623}
}
\email{mac9908@rit.edu}
%
%
\author{Naveen Sharma}
\affiliation{%
 \institution{Rochester Institute of Technology}
 \city{Rochester} 
 \state{NY} 
 \postcode{14623}
}
\email{nxsvse@rit.edu}

\renewcommand{\shortauthors}{Chowdhury et. al.}

\begin{abstract}
Vacant lots are neglected properties in a city that lead to environmental hazards and poor standard of living for the community. Thus, reclaiming vacant lots and putting them to productive use is an important consideration for many cities. Given a large number of vacant lots and resource constraints for conversion, two key questions for a city are
\begin{enumerate*}[label=(\arabic*)]
\item whether to convert a vacant lot or not; and
\item what to convert a vacant lot as.
\end{enumerate*}
We seek to provide computational support to answer these questions. To this end, we identify the determinants of a vacant lot conversion and build a recommender based on those determinants.
We evaluate our models on real-world vacant lot datasets from the US cities of Philadelphia, PA and Baltimore, MD. Our results indicate that our recommender yields mean F-measures of
\begin{enumerate*}[label=(\arabic*)]
\item 90\% in predicting whether a vacant lot should be converted or not within a single city, 
\item 91\% in predicting what a vacant lot should be converted to, within a single city and,
\item 85\% in predicting whether a vacant lot should be converted or not across two cities.
\end{enumerate*}
\end{abstract}

%
%
 




\maketitle

\section{Introduction} \label{sec:intro}

A vacant lot is an abandoned property that has no buildings on it. In many cases, there were buildings on vacant lots but they were abandoned due to suburbanization, loss of industrial base, and other such reasons. As the abandoned buildings fell into disrepair and caused safety concerns, they were demolished \cite{vacantlotproblem}, and the corresponding properties became vacant lots.

Vacant lots are a major concern for cities because they tend to
\begin{enumerate*}[label=(\arabic*)]
\item be subject to litter and dumping of other solid wastes, including hazardous waste such as lead and asbestos; 
\item shelter criminal activities; and 
\item depress property values in their neighborhood \cite{bucchianeri2012valuing}.
\end{enumerate*}
Overall, vacant lots result in a poorer standard of living for urban communities. Thus, reclaiming vacant lots and putting them into productive is a primary concern for many communities \cite{kremer2013social}. 

A primary strategy cities employ to deal with vacant lots is to fill the empty spaces with attractive and productive activities. For example, the vacant lots of Baltimore have been converted to community gardens, urban farms, and community managed open spaces. In some cases, filling the vacant lots not only attracts people, but also increases the standard of living of the community \cite{pyle2002eden}. However, it may not be feasible to convert all vacant lots in a city because of resource constraints. In such cases, cities need to decide
\begin{enumerate*}[label=(\arabic*)]
\item which vacant lots, if converted, provide the most benefits; and 
\item what those vacant lots should be converted to.
\end{enumerate*}


Although much research has been done in analyzing the impact of converting vacant lots \cite{bucchianeri2012valuing, heckert2012economic, garvin2013greening}, none of the existing works focus on optimally choosing vacant lots for conversion. In practice, urban planners study the effect of converting a vacant lot and make decisions based on it, and community members make a selection recommended by the urban planners. In this process, cities reinvent the wheel every time, making these decisions from scratch. In contrast, modeling vacant lot conversion as a data-driven problem can minimize the redundancies in selecting lots to convert. 

Programs such as ``Grounded In Philly'' in Philadelphia \cite{groundedinphilly} and ``Adopt-A-Lot'' in Baltimore \cite{adoptalotbalti} provide publicly available data on locations of vacant lots, including those that have been converted. Given such historical data, i.e., data about which lots have been converted in a city and to what, our objective is to answer the following two research questions: 
\begin{description}
\item[RQ1] Can we predict whether a vacant lot should be converted?
\item[RQ2] Can we predict what should a vacant lot be converted to?
\end{description} 

To answer these research questions, we propose {\rec}, a recommender for urban vacant lot conversion. {\rec} learns to recommend optimal vacant lots for conversion based on historical data about vacant lot conversion and six types of features (determinants) that determine whether a vacant lot should be converted and what it should be converted to. {\rec} provides urban planners with preliminary predictions on which vacant lots should be converted, and analyze why these vacant lots were chosen. This provides a starting point for urban planners to focus limited resources to prioritize certain vacant lots. 

\subsection*{Contributions}
\begin{description}
\item [Determinants:] We identify six determinants of vacant lot conversion:
\begin{enumerate*}
\item utility from public services and infrastructure,
\item access to vacant lot,
\item neighborhood property value indicator,
\item vacant lot density,
\item crime density, and
\item zoning policies.
\end{enumerate*}
These determinants serve as features in answering the two prediction questions we ask above.
\item [Datasets:] We build two datasets based on the cities of Baltimore and Philadelphia, which include the features above for vacant lots and vacant lots that have been converted.
\item [Recommender:] We propose {\rec}, a recommender that predicts whether a vacant lot should be converted or not, and to what, if the lot is to be converted, based on the datasets above.
\end{description}

\subsubsection*{Organization}
Section~\ref{sec:rel_work} reviews related works. Section~\ref{sec:attributes} describes the proposed vacant lot model and details the determinants of a vacant lot conversion. Section~\ref{sec:datasets} outlines the datasets that we built for this research. Section~\ref{sec:experiments} describes the experiments we performed on {\rec}, and Section~\ref{sec:result} discusses the results from those experiments. Section~\ref{sec:future} discusses some of the applications of this research and future work, and Section~\ref{sec:conclusion} provides conclusions.	

\section{Related Work} \label{sec:rel_work}

The concept of vacant lots and possible solutions are studied extensively well in existing literature. Accordino {\etal} \cite{vacantlotproblem} describes how the existence of vacant lots cause concern for the community, and provides an overview of how they are solved in various cities. Accordino \cite{vacantlotproblem}  concludes that the solution to such problems not only happen from an urban planning side, but also from the neighborhood community as well. However, the study is focused toward cities as a whole and does not take into account whether each problem tackled resulted in success for those vacant lots.

Branas {\etal} \cite{vacantlotgreening} looks at the effects of vacant lot conversion where estimates showed that vacant lot greening was associated with consistent reductions in gun assaults across all four sections of the city. Vacant lot greening also resulted in consistent reductions in vandalism in one section of the city. Regression-adjusted estimates also showed that vacant lot greening was associated with residents reporting less stress and more exercise in select sections of the city. Once greened, vacant lots may reduce certain crimes and promote some aspects of health. 

Furthermore Kremer \cite{kremer2013social} suggests that by assessing vacant lot uses, ecological characteristics and the social characteristics of neighborhoods in which vacant lots are located, planners may be able to more effectively address urban land vacancy while supporting urban sustainability and resilience. Automating such analysis to reduce the burden on urban planners is one of the primary motivation for the development of {\rec}.

Recommender systems have been used extensively, and their use in different types of data are highlighted in Capdevila {\etal} \cite{geosrs} for geolocation based data and in Ramesh {\etal} \cite{recommenderSocialMedia} for social media data. These papers, however, act more as a proof of concept rather than a solution to an existing problem.

Tayebi {\etal} \cite{tayebi2011crimewalker} presented a novel approach to crime suspect recommendation based on partial
knowledge of offenders involved in a crime incident and a known co-offending network.

Ruining {\etal} \cite{he2016vista} built a large-scale recommender systems to model the dynamics of a vibrant digital art community, Behance, consisting of tens of millions of interactions (clicks and `appreciates') of users toward digital art.

The recommender system to be developed needs to estimate the present utility of the vacant lot along with the future utility after a conversion. The concept of utility for urban infrastructure, including vacant lots, was introduced by Meidar-Alfi \cite{urbanutility}, along with a detailed case study analysis. Her paper proposes the relationship between benefits of an infrastructure to the distance from the infrastructure. The author analyzes the inverse relationship for utility with distance from the infrastructure. Our research utilizes this theory, and models the benefit a vacant lot may receive in terms of determinants that include distances from parks, libraries and schools.

\section{The Vacant Lot Model}\label{sec:attributes}
This section proposes a unified formal model of describing a vacant lot in terms of attributes related to its neighborhood. This model assumes that each vacant lot has a set of features that determine if the vacant lot should be converted. This model also assumes that those features will  determine what a vacant lot should be converted to. 


Our vacant lot model consists of six types of features. The following subsections provide a detailed description on each of the features along with the rationale behind their selection. We also detail how the features are represented in our vacant lot model.

\subsection{Utility from Public Services and Infrastructure} 
A primary indicator of the wellness of any land parcel is the utility it receives from its closest public infrastructure. As described in Meidar-Alfi \cite{urbanutility}, distance from these infrastructure is inversely proportional to the utility a land parcel may receive. 

Public infrastructure common to all cities are public libraries, schools and parks. In order to get an indication of utilities provided by these facilities, for each vacant lot we calculated the distance in meters to the closest library, park and school. An increasing distance from the closest public infrastructure would result in a decreased utility received from those infrastructures. After the necessary calculations, we get three features \textit{distance to school}, \textit{distance to park}, \textit{distance to library}.

\subsection{Access to Vacant Lots}
Once a vacant lot is converted, the benefit it provides will be dependent on the ease of access it has. A study by Wachter {\etal} \cite{vacantlotaccess} analyzes the effect of public transit on vacant land management, and suggests that public transit may be a determinant for the impact a vacant lot has upon conversion. Similar to our measure of public infrastructure utility, we also measure the distance from each vacant lot to the nearest public transit stop for another feature, \textit{distance to transit stop}.

\subsection{Neighborhood Property Value} \label{sec:land_value} 
There is a significant amount of work in current literature that focuses on the impact of vacant lots on neighborhood property values. Most of the work done focuses on a hedonic or spatial difference-in-difference analysis of the impacts \cite{heckert2012economic, bucchianeri2012valuing, vacantlotgreening}. We utilize concepts from both these approaches in estimating how a vacant lot or community garden affects the neighborhood property values at present time.

\begin{figure}[!htb]
\includegraphics[width=\columnwidth]{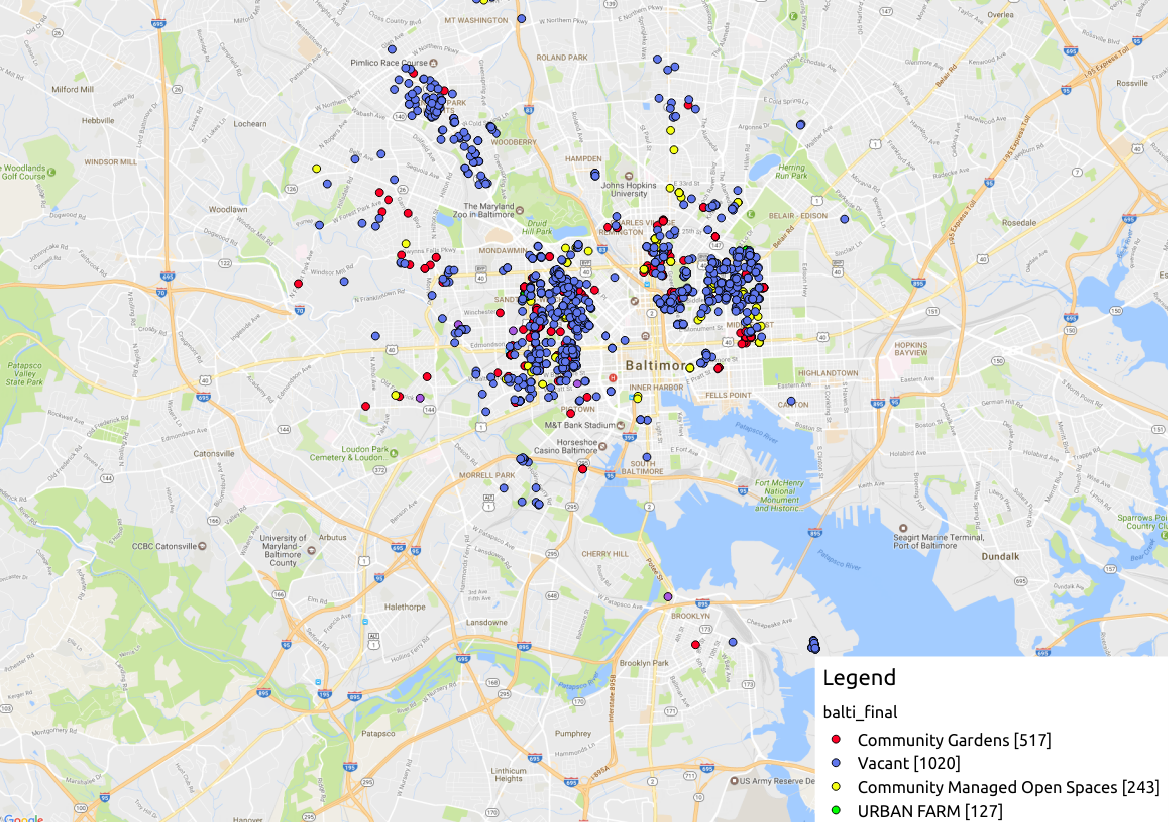}
\centering
\caption{Vacant lots \& community gardens in Baltimore}
\label{fig:balti_vacant}
\end{figure}

We calculate the mean property value in a quarter mile radius for each vacant lot for two points in time. For the purpose of our research, we chose property assessment data for the years 2015 and 2014 due to the availability of recent data. The difference between the two points provides a simplified estimate of the trend in property values, and may indicate how the immediate surrounding area is affected by the existence of vacant lots. A better estimate would have been to compare property values before a vacant lot was converted to a community garden, but due to the unavailability of such data a much more simplified estimate is used.

Another indication of the status of the area a vacant lot is situated in is the median property values for that area. In most cases, vacant lots situated in a higher value market may provide better benefits upon conversion than those in more distressed market, and as such urban planners may take greater initiative to convert them. Furthermore, the existence of vacant lots will negatively affect the overall market value of a neighborhood in the long term \cite{vacantlotgreening}.

\subsection{Vacant Lot Density} \label{sec:vacant_density}
While deciding determinants of a vacant lot conversion, we needed to consider spatial characteristics of each vacant lot. To that end, for each vacant lot we needed an indication of how many vacant lots are in its surrounding. As a result, vacant lot density is calculated. Vacant lot density for each individual vacant lot is defined as the number of vacant lots in a quarter mile radius from the respective vacant lot. 

\subsection{Crime} \label{crime_density}
Garvin {\etal} \cite{garvin2013greening} performed a randomised controlled trial of vacant lot conversion to test its impact on crime in the surrounding area. Results indicated that there is an association between conversion and crime, and vacant lot conversion decreases crime incident. Due to the extent of the impact on crime, we decided to utilize crime as a determinant of our vacant lot model.

We use crime density as a feature in our model, and crime density is calculated using the number of criminal incidents that were reported in a quarter mile radius from each vacant lot in one year.
\begin{figure}[t]
\includegraphics[width=\columnwidth]{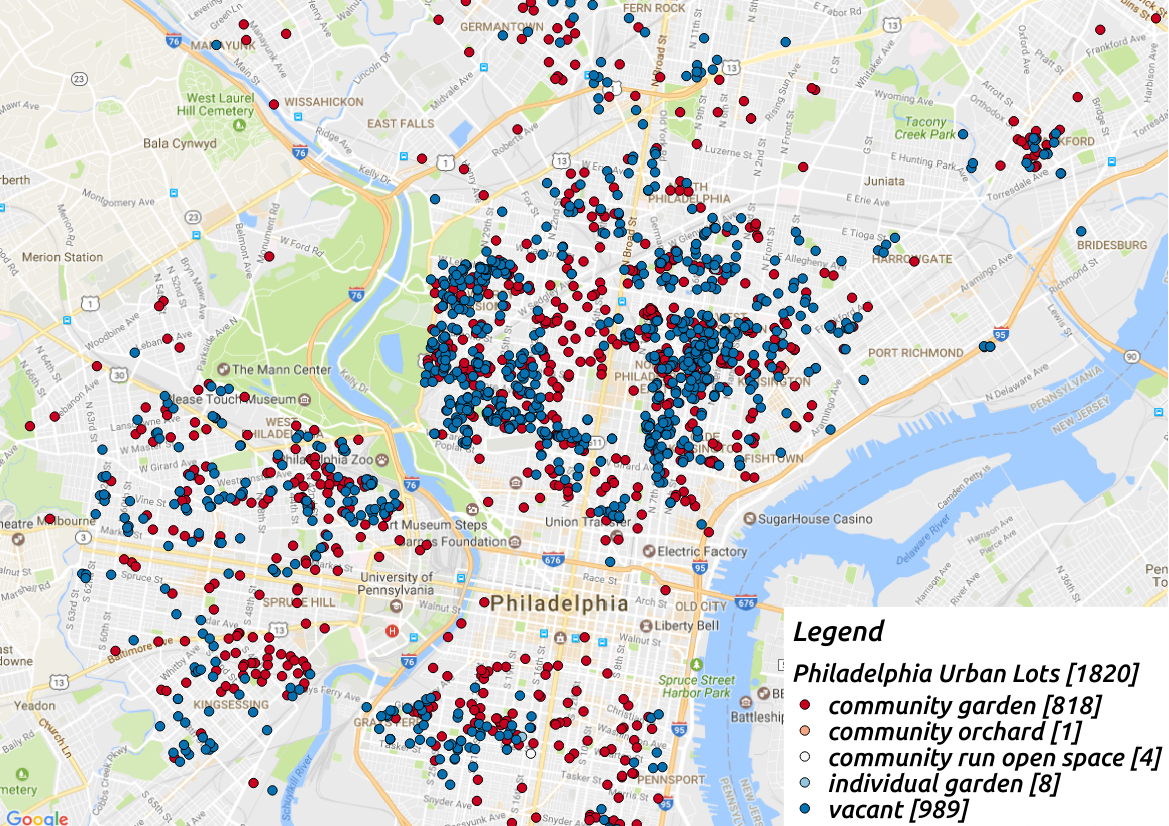}
\centering
\caption{Vacant lots \& community gardens in Philadelphia}
\label{fig:philly_vacant}
\end{figure}

\begin{table}[!htb]
\centering
\caption{Determinants of a vacant lot conversion}
\label{tab:summary_determinants}
\begin{tabular}{ll}
\toprule
\textbf{Variable} & \textbf{Description} \\ 
\midrule
libDist (numeric)            & Distance from the closest library \\
parkDist (numeric)           & Distance from the closest park \\
schoolDist (numeric)         & Distance from the closest school \\
transitDist (numeric)        & Distance from the closest transit stop \\
priceDiff (numeric)          & Difference in mean property value \\
vacantDensity (numeric)      & Density of vacant lots \\
crimeDensity (numeric)       & Density of crime incidents \\
zone (categorical)           & Zoning policy for vacant lot \\ 
\bottomrule
\end{tabular}%
\end{table}

\subsection{Zoning Policies}
Zoning is the process of dividing land in a municipality into zones (e.g. residential, industrial) in which certain land uses are permitted or prohibited. Thus, zoning is a technique of land-use planning as a tool of urban planning used by local governments in most developed countries.

Every city divides its land into zones with a specific purpose. Each zone defines what can and cannot be built upon the vacant lot, or whether it can be converted as well \cite{babcock1985zoning}. Since zoning policies dictate the development of vacant lots, we used it as an attribute for our vacant lot model. Zoning policies are categorical variables that are either residential, industrial, business or special purpose.  

All the determinants of our vacant lot model is summarized in Table \ref{tab:summary_determinants}.

\begin{figure}[!htb]

\includegraphics[width=\columnwidth]{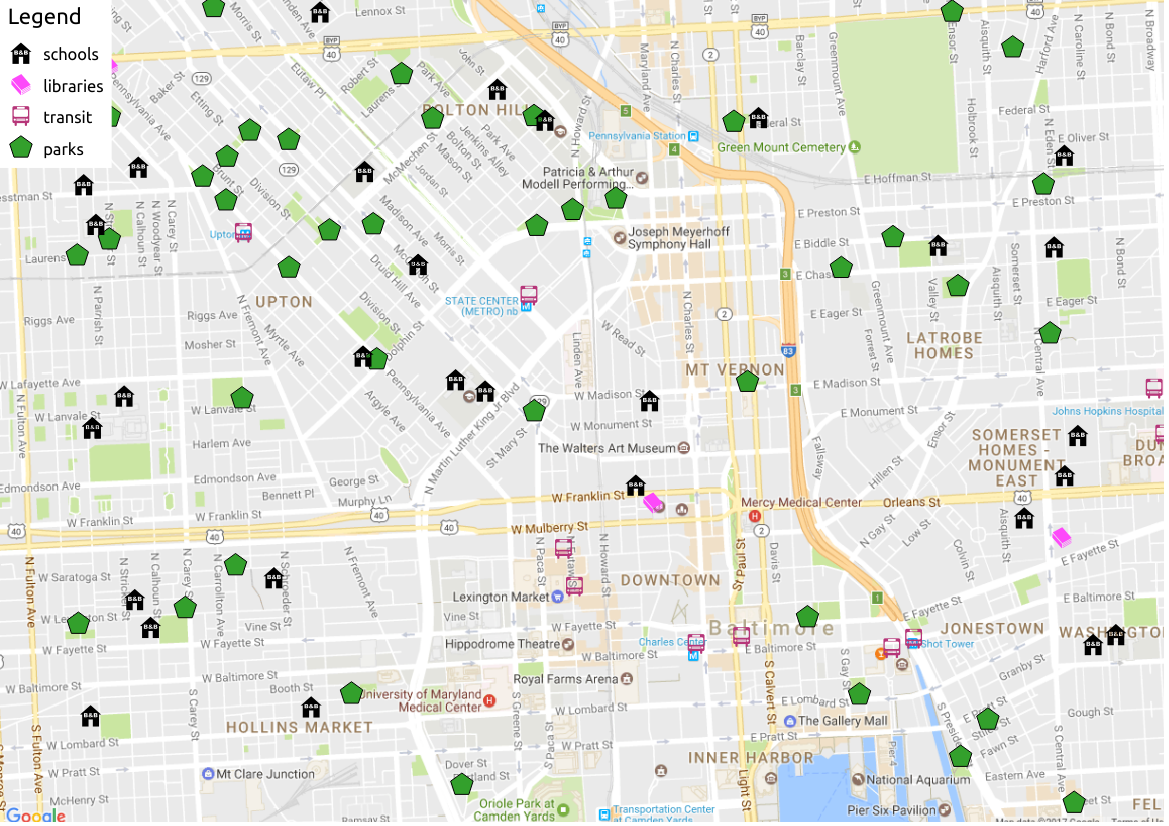}
\centering
\caption{Public infrastructures in Baltimore}
\label{fig:balti_public}
\end{figure}


\section{Datasets} \label{sec:datasets}

This section discusses the datasets used to build the proposed vacant lot model for our experiments. The two primary datasets this research utilizes are from the cities of Baltimore, MD and Philadelphia, PA. The primary reason for the choice of these two cities was the availability of sufficient quantifiable data on the status of current vacant lots, and also their determinants. In order to build this dataset, we utilized several different data sources. 

\begin{table}[!htb]
\centering
\caption{Vacant lot distribution in the datasets}
\label{tab:dataset_distribution}
\begin{tabular}{lrrr}
\toprule
\textbf{City} & \textbf{Total} & \textbf{Vacant} & \textbf{Converted}\\
\midrule
      Baltimore & 1907 & 1020 & 887 \\
      Philadelphia & 1820 & 989 & 831 \\
\bottomrule      
\end{tabular}
\end{table}

Vacant lot data for Baltimore was acquired from the Adopt-A-Lot program \cite{adoptalotbalti}, which provides location and status of the vacant lots in the city of Baltimore. For the city of Philadelphia, the vacant lot data was acquired from the Grounded In Philly program \cite{groundedinphilly} which provided similar data. The distribution of the vacant lots and converted vacant lots are given in Table~\ref{tab:dataset_distribution}. Figure~\ref{fig:philly_vacant} shows the distribution of vacant lots and community gardens in our dataset for the city of Philadelphia.   

\begin{figure}[!htb]
\includegraphics[width=\columnwidth]{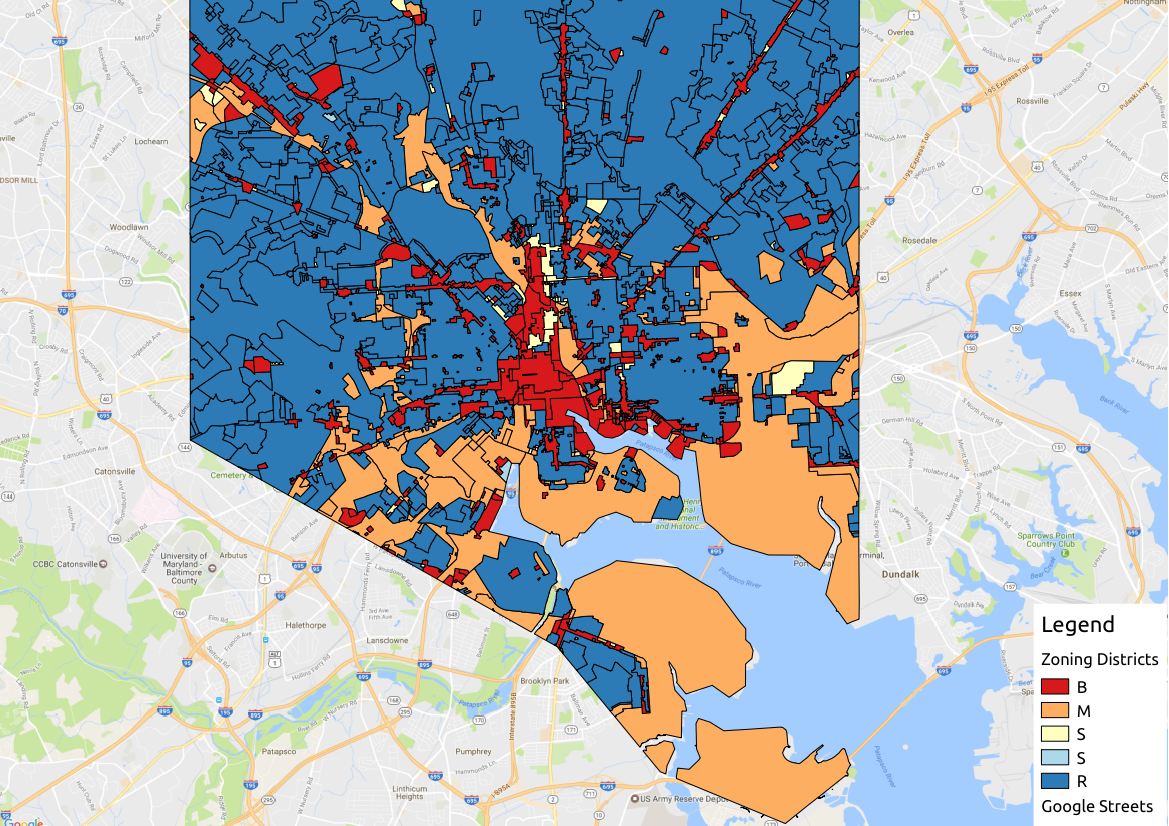}
\centering
\caption{Zoning for the City of Baltimore}
\label{fig:balti_zoning}
\end{figure}

\begin{figure}[!htb]
\includegraphics[width=\columnwidth]{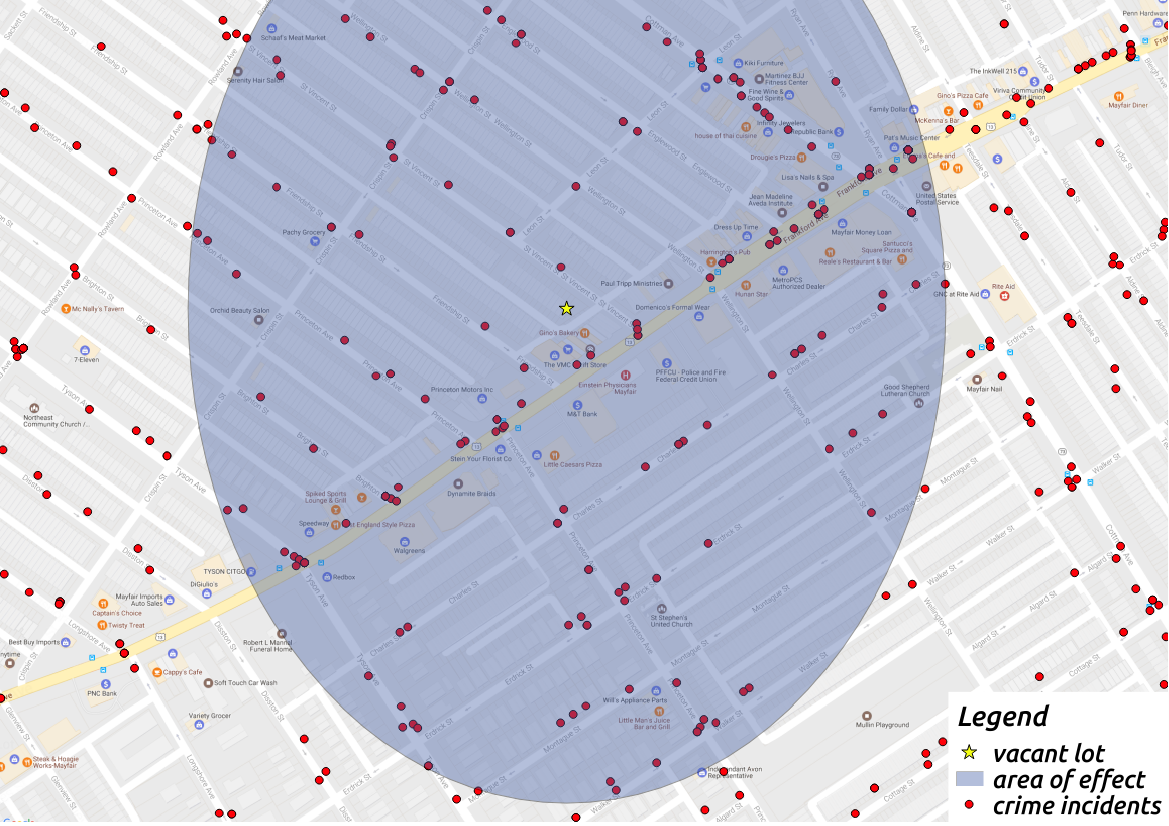}
\centering
\caption{Crime density from the Philadelphia dataset}
\label{fig:philly_crime}
\end{figure}

Locational data on public service infrastructures  such as libraries, parks and schools, along with public transit stops for Baltimore was acquired from Open Baltimore \cite{baltigis}, while for Philadelphia it was acquired from OpenDataPhilly \cite{phillygis}.   
Publicly available crime incident reports for the year 2015 was acquired for both Baltimore and Philadelphia from \cite{baltigis} and \cite{phillygis} respectively. Then, we used the vacant lot data to calculate crime density for a quarter mile radius of each vacant lot as shown in Figure~\ref{fig:philly_crime}. Zoning districts were also acquired from \cite{baltigis} and \cite{phillygis}, and vacant lot data were spatially joined with the predetermined zoning districts to categorize which district each vacant lot belonged to. An example for the Baltimore zoning district is given in Figure \ref{fig:balti_zoning}.

\begin{figure}[!htb]

\includegraphics[width=\columnwidth]{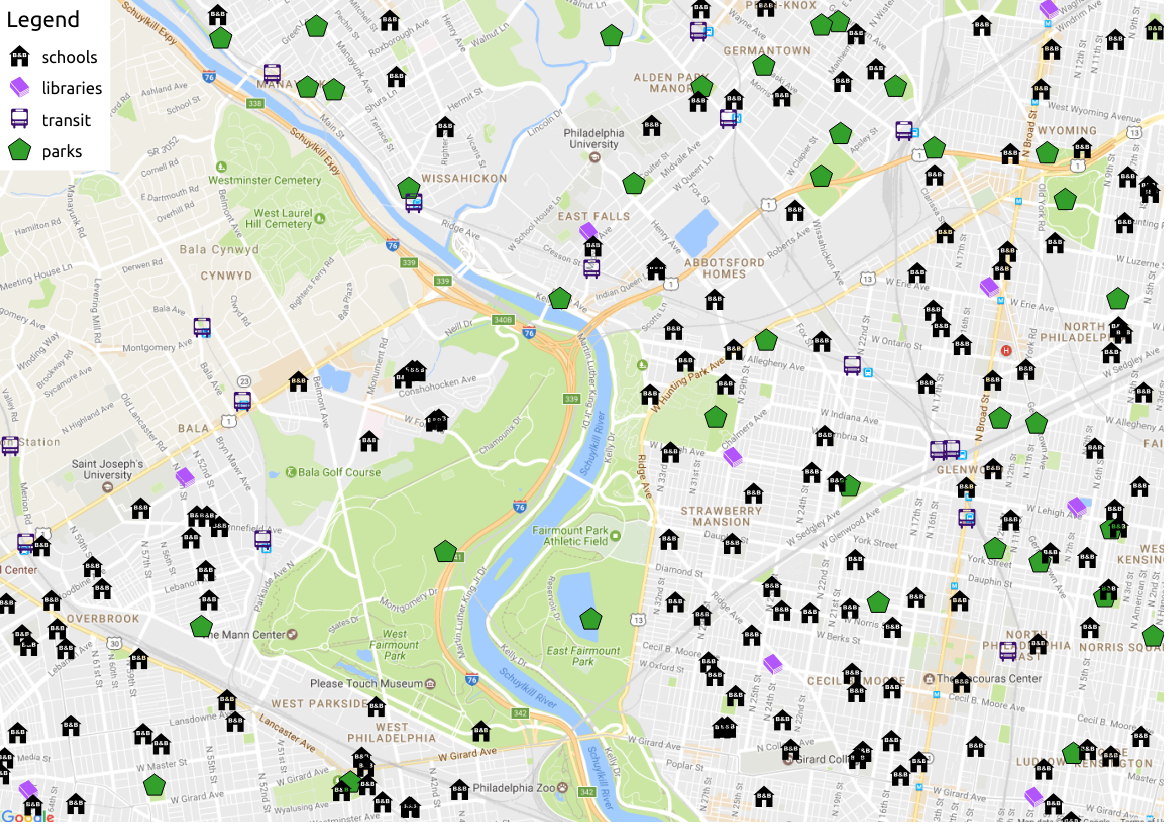}
\centering
\caption{Public infrastructures in Philadelphia}
\label{fig:philly_public}
\end{figure}

We utilized yearly property assessment datasets from Baltimore \cite{adoptalotbalti} and Philadelphia \cite{phillygis}, and collected property prices for the year 2014 and 2015. For each vacant lot, we calculated the mean price of properties within a quarter mile radius of the particular vacant lot for both 2015 and 2014. The difference between the two years indicated our price difference variable.

The dataset proportion is given in Table~\ref{tab:dataset_distribution}, with the size of each dataset, proportion of vacant lots and their conversions. After all the data was collected, a final dataset was built to represent the vacant lot model.

\section{Experiments}\label{sec:experiments}
This section describes the methodology and process we followed in experimenting with the generated dataset described in Section~\ref{sec:datasets}. The objective for these experiments were to build multiple prediction models on different combinations of our predetermined independent variables, evaluate the accuracy of our classifiers and select the best one that will be used in {\rec}.

\begin{table}[!htb]
\centering
\caption{Performance of feature subsets for a Random Forest classifier on Baltimore}
\label{tab:features}
\resizebox{\columnwidth}{!}{%
\begin{tabular}{@{}lllll@{}}
\toprule
\multirow{2}{*}{Feature}    & \multicolumn{3}{l}{Convert} & Overall  \\ \cmidrule(l){2-5} 
                            & Precision & Recall & F1   & Accuracy \\
\midrule                            
libDist+parkDist+schoolDist & 0.89      & 0.81   & 0.85 & 0.89     \\
transitDist+zone        & 0.66      & 0.62   & 0.64 & 0.75     \\
vacantDensity+crimeDensity  & 0.80      & 0.75   & 0.77 & 0.84     \\ 
\bottomrule
\end{tabular}%
}
\end{table}

We first analyze how each of the features can be used to select a vacant lot. Since we are utilizing the features to get a general idea of how they determine vacant lot conversion, we only performed this experiment on Baltimore. The dataset was split into a training \& testing set consisting 60\% of the data, while a validation set containing 40\% of the data was also created. We built a simple Random Forest classifier on a subset of features using the training set and evaluated their results based on our validation set. The results are given in Table~\ref{tab:features}, which indicates that features such as the public infrastructure distances, vacant lot density and crime density provide an accurate model to predict if a vacant lot should be converted. However, even if transit distance and zoning have a weaker association, their effect is still better than random, indicating their usefulness. As a result, we incorporated all the proposed features for our vacant lot model in implementing {\rec}.

We also had to consider the effectiveness of different classifiers incorporated in {\rec}. As a result, we performed our experiments on the following classifiers:
\begin{description}
\item[Random Forest] Random forests are an ensemble learning algorithm used for classification problems. Each random forest consists of multiple decision trees that are constructed at training time and classifying based on the each of the attributes in the data. Random forest was developed to tackle the problem of decision trees overfitting to the training data \cite{kanevski2009machine}.
\item[k-NN] k-NN is categorized as a lazy learner, and falls under the class of instance based learners. It utilizes similarity between objects and an unknown object is classified by a majority vote of its most similar objects. Furthermore, k-NN does not build a model and only approximates a prediction upon receiving an unknown instance.
\item[MLP] A multilayer perceptron (MLP) is a feed-forward artificial neural network. It consists of input nodes, multiple layers of hidden nodes. Each layer is connected to the next layer, and the network itself is represented as a directed graph. Each node is responsible for processing the input data with the help of an activation function. With each iteration, the network is trained with a backpropagation algorithm that enables weights to be updated with each training instance coming in, to decrease the error of predictions made.
\item[Naive Bayes] Naive Bayes is a classification algorithm based on Bayes’ Theorem that assumes that each of the independent variables are independent of one another. It's the simplest form of a Bayesian classifier, and it's strength with categorical variables suit the design of the vacant lot problem as well.
\item[SVM] SVM is a classification technique that constructs linear separating hyperplanes in high-dimensional vector space to separate data points based on their features. The purpose of an SVM is to maximize the separation of the data points from these hyperplanes in order to increase confidence of the classification.
\end{description}

The following sections describe the experiments we performed on {\rec} in multiple scenarios utilizing different classification algorithms. 

\subsection{Vacant Lot Selection in a City}

At first, we build a prediction model for {\rec} focused on a single city and study which vacant lots should be converted within a single city. For our experiments we created five prediction models each for the cities of Baltimore and Philadelphia, with hyperparameters tuned to optimize results on the training set for each city. 

In building those classifiers, our training set was a random sample of 60\% of the primary dataset. The remaining 40\% of the datasets were used to evaluate the accuracy of our model, and ensure the models provide valid and sane predictions. The classifiers were trained on the training set using a 5-fold cross validation to prevent the models from over-fitting to the training data. The results are discussed in detail in Section~\ref{sec:result}.

\subsection{Vacant Lot Conversion in A City}

Once we have experimented on selecting a vacant lot within a city using {\rec}, we evaluate our model further by studying if {\rec} can predict what a vacant lot should be converted to based on historical datasets in a single city. However, only the dataset for Baltimore includes sufficient historical data on what a vacant lot has been converted to. As a result, we focus our experiments on Baltimore dataset. 

In the dataset, there are three classes of conversions for Baltimore:
\begin{enumerate*}
\item Community Gardens,
\item Urban Farms, and
\item Qualified Community Managed Open Space (QCMOS).
\end{enumerate*}

We create three class classifiers for Baltimore using our training set and evaluate it against a validation set. For this particular experiment, our model is tested twice. Once with the assumption that all vacant lots are suitable for conversion, and once with no such assumption at all. The results are discussed in detail in Section~\ref{sec:result}.

\subsection{Vacant Lot Selection Between Two Cities}

After analyzing the prediction models for each city, we picked the best classification algorithm and incorporated them into {\rec} to select vacant lots across cities. We trained our classifiers on Baltimore and tested it on Philadelphia, and vice versa. However, the two test cities in our experiments have different classes that vacant lots were converted to. For example, in Baltimore vacant lots were converted to Qualified Community Managed Open Space (QCMOS), urban farms or simply adopted to a community garden. But in Philadelphia our dataset only consists of community garden conversions. As a result, both dependent variables were changed to a binary class indicating whether a vacant lot has been converted or not. 

After the dataset has been updated, similar experiments as described in the previous section was carried out. However, in this case we utilized Baltimore as our training set and validated the results with the datasets from Philadelphia, and vice versa. We also included multiple combinations of training sets, such as utilizing 20\% of Baltimore and 100\% of Philadelphia datasets and validating against the remaining 80\% of the Baltimore dataset. This was to ensure that our models captured at least some parts of target cities.

\section{Results}\label{sec:result}

This section discusses the results from the experiments described in the Section~\ref{sec:experiments}. We started by creating three separate simplified Random Forest classifiers for a subset of features to analyze how they interact with our target variable. As can be seen in Table~\ref{tab:features}, public utility such as distance from library, park etc. provides a strong indication of the vacant lot conversions, while transit distance and zoning comparatively has weaker association. However, their prediction accuracy is still significantly better than random, and hence their contribution cannot be ignored.

\begin{table}[!htb]
\centering
\caption{Hyperparameter tuning for Random Forest classifier on Baltimore}
\label{tab:hyper_rf}
\begin{tabular}{ccc}
  \toprule
 mtry & Accuracy & AccuracySD \\ 
  \midrule
2 & 0.81 &  0.01 \\ 
  8 & 0.83 & 0.01 \\ 
  14 & 0.83 & 0.02 \\ 
   \bottomrule
\end{tabular}
\end{table}

\begin{table}[!htb]
\centering
\caption{Hyperparameter tuning for MLP classifier on Baltimore}
\label{tab:4.2a}
\begin{tabular}{ccc}
  \hline
 size & Accuracy & AccuracySD \\ 
  \midrule
3 & 0.59 & 0.04 \\ 
   5 & 0.63 & 0.04\\ 
   8 & 0.66 & 0.03 \\ 
   10 & 0.69 & 0.01 \\ 
   \hline
\end{tabular}
\end{table}

\begin{table}[!htb]
\centering
\caption{Summary of vacant lot selection for Baltimore}
\label{tab:balti_binary}]
\begin{tabular}{@{}llll@{}}
\toprule
\multirow{2}{*}{Classifier} & \multicolumn{3}{c}{Adopt} \\ \cmidrule(l){2-4} 
 & Precision & Recall & F1 \\ 
 \midrule
Random Forest & 0.88 & 0.85 & 0.86 \\
k-NN & \textbf{0.89} & \textbf{0.88} & \textbf{0.89} \\
SVM & 0.67 & 0.52 & 0.59 \\
MLP & 0.61 & 0.73 & 0.67 \\
Naive Bayes & 0.91 & 0.50 & 0.65 \\ \bottomrule
\end{tabular}%
\end{table}

\begin{table}[!htb]
\centering
\caption{Summary of results for the Philadelphia dataset}
\label{tab:philly_summary_binary}
\begin{tabular}{llll}
\toprule
\multirow{2}{*}{Classifier} & \multicolumn{3}{c}{Adopt}\\ \cmidrule(l){2-4} 
                            & Precision     & Recall        & F1            \\ \midrule
Random Forest               & \textbf{0.90} & \textbf{0.93} & \textbf{0.92} \\
k-NN                        & 0.85          & 0.84          & 0.84          \\
SVM                         & 0.67          & 0.72          & 0.69          \\
MLP                         & 0.63          & 0.79          & 0.70          \\
Naive Bayes                 & 0.76          & 0.72          & 0.74          \\ \bottomrule
\end{tabular}
\end{table}

We utilized two datasets from the cities of Baltimore and Philadelphia to perform three major experiments on analyzing vacant lot selection within a city, vacant lot selection across cities, and vacant lot conversion. The details of the results are discussed in the following sections.

\subsection{Vacant Lot Selection in a City}

The training set for both Baltimore and Philadelphia consisted of a random sample of 60\% of the data. The remaining 40\% was used to validate our prediction model. The vacant lot data for both cities consisted of binary classes, Adopt or Available. Adopt indicated that the vacant lot should be converted, while Available indicated that it should not.

\begin{figure*}[!htb]
\centering
\subfigure[Observed]{\label{fig:balti_obs}\includegraphics[width=\columnwidth]{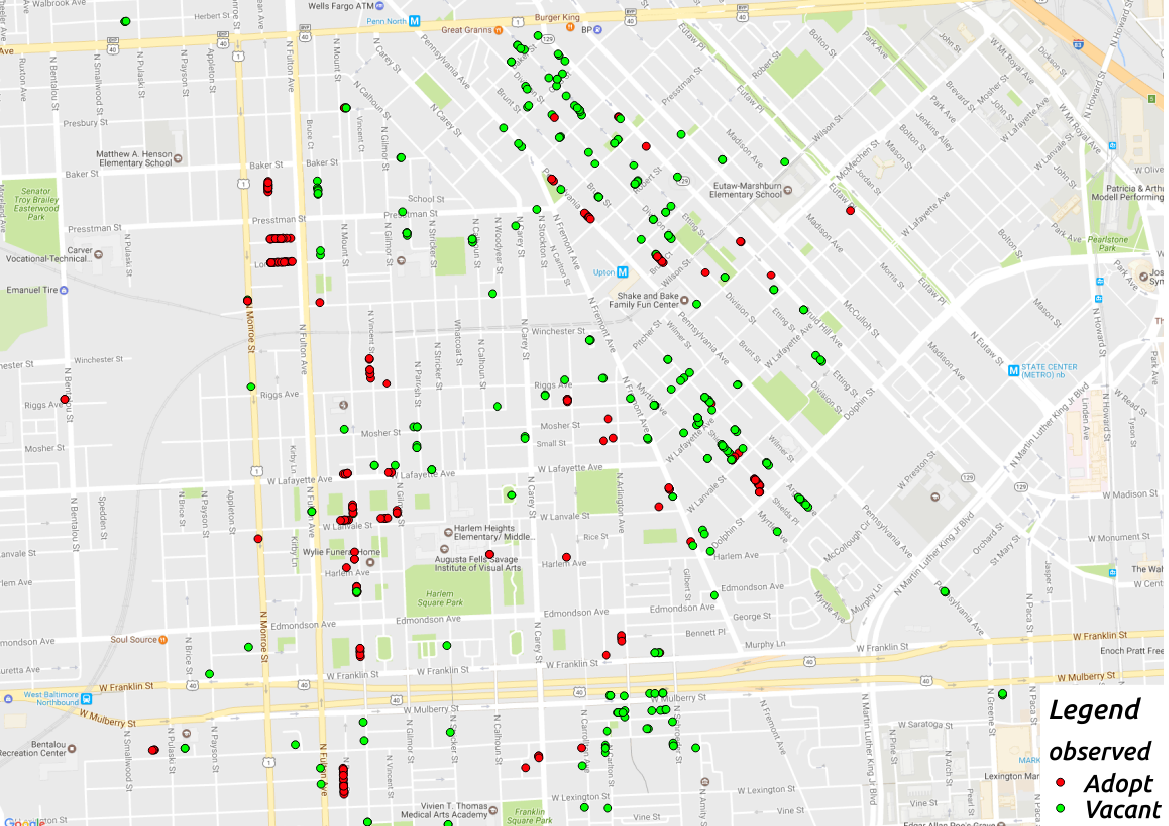}}
\subfigure[Predicted]{\label{fig:balti_pred}\includegraphics[width=\columnwidth]{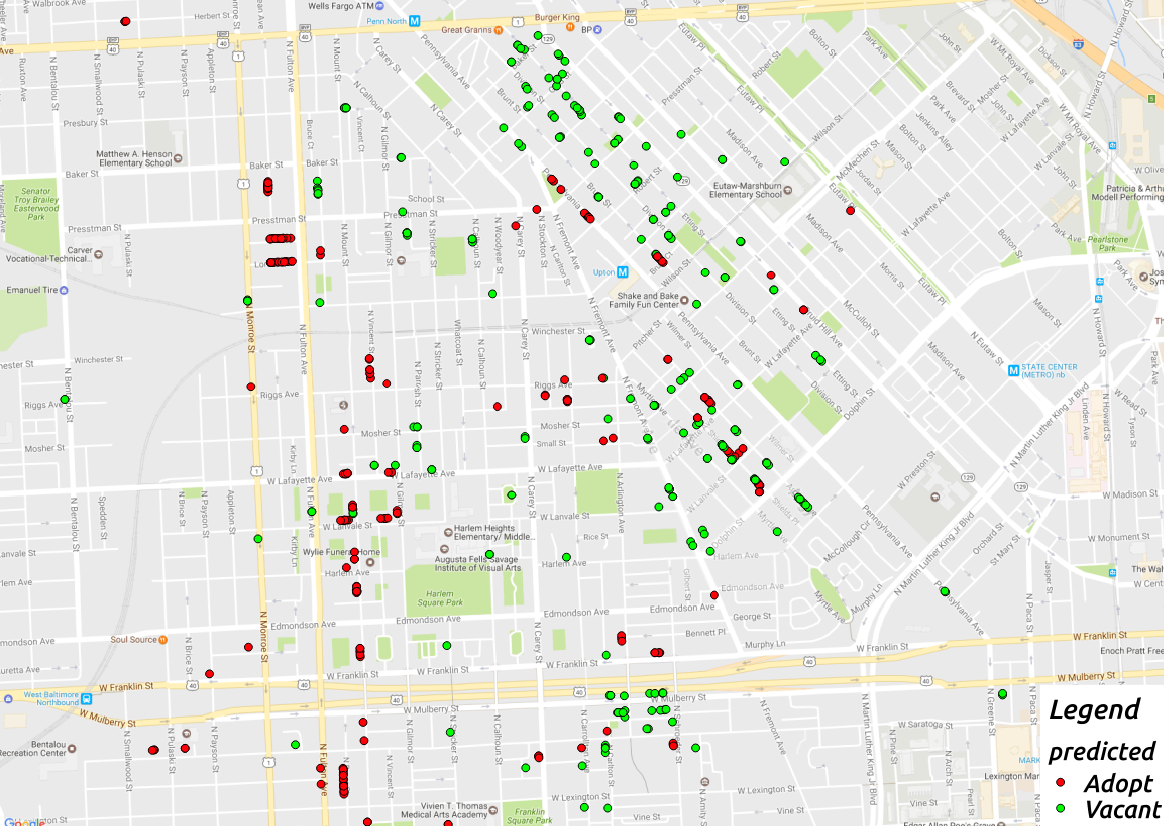}}
\caption{Observed \& Predicted values for Vacant lot conversion}
\label{fig:balti_test}
\end{figure*}

The results for the Baltimore dataset for each classifiers used are:
\begin{description}
\item[Random Forest] The first experiment focused on using a Random Forest as the primary classifier for {\rec}. We tuned our classifier using the number of trees in our Random Forest as a parameter $mtry$. As given in Table~\ref{tab:hyper_rf}, the highest accuracy was provided with 8 decision trees in our Random Forest. The precision, recall and F1 score of the Random Forest classifier is sufficiently high and indicates that within a single city Random Forest can be used as the primary classifier for {\rec}.

\item[MLP] The second experiment focused on the use of a Multilayer Perceptron (MLP) to build our prediction model. After performing hyperparameter tuning, the best results were obtained using an MLP with 10 hidden nodes as indicated by the variable $size$ in Table~\ref{tab:4.2a}. Compared to Random Forest, MLP performs poorly in our Baltimore dataset, with a particularly low precision. 

\item[Naive Bayes] The third experiment focused on the use of a Naive Bayes classifier to build our prediction model. In terms of performance, Naive Bayes performs slightly better than MLP, but does not perform as well as a Random Forest.
\item[k-NN] The fourth experiment focused on the use of a k-Nearest Neighbor classifier to build our prediction model. The hyper-parameter tuned for this particular classifier was the number of neighbors to consider for similarity of a vacant lot. The most optimal result was obtained for $k=1$. k-NN outperforms Random Forest for the Baltimore dataset in terms of precision, recall and accuracy. Due to the use of similarity between vacant lots k-NN does mimic the process an urban planner might take in choosing to convert a vacant lot, and as a result is better able to capture the pattern.
\item[SVM] The fifth experiment focused on the use of an SVM classifier to build our prediction model. In terms of performance, SVM performs similar to MLP, but does not perform as well as a Random Forest and k-NN.
\end{description}

The final results from our experiments are given in Table~\ref{tab:balti_binary}. As can be seen from the table, k-NN and Random Forest provides the best precision and recall. Naive Bayes performs well in terms of precision but has a significantly lower recall. MLP and SVM only perform well to a certain extent.

The dataset for Philadelphia was also split into a training \& testing set consisting of a random sample of 60\% of the data, while the remaining 40\% were left for validation purposes. Each of the experiments that were performed on Baltimore dataset was performed again on the Philadelphia dataset. 

The overall results are given in Table~\ref{tab:philly_summary_binary}. For the Philadelphia dataset, there were only two classes with vacant lots as Available and community gardens as Adopt. Similar to the results in Baltimore, Random Forest and k-NN performs the best in predicting which vacant lots should be converted.

\subsection{Vacant Lot Conversion in A City}

\begin{table*}[t]
\centering
\caption{Summary of results for predicting what the vacant lots should be converted to}
\label{tab:balti_threeclass}
\resizebox{\textwidth}{!}{%
\begin{tabular}{@{}lcccccccccccc@{}}
\toprule
\multirow{2}{*}{Classifier} & \multicolumn{3}{c}{ADOPTED} & \multicolumn{3}{c}{QCMOS} & \multicolumn{3}{c}{URBAN FARM} & \multicolumn{3}{c}{Overall (Mean)}  \\ 
\cmidrule(lr){2-4} \cmidrule(lr){5-7} \cmidrule(lr){5-7} \cmidrule(lr){8-10} \cmidrule(lr){11-13}
                            & Precision  & Recall  & F1   & Precision & Recall & F1   & Precision   & Recall   & F1    & Precision & Recall & F1 \\ 
\midrule
Random Forest               & 0.90       & 0.94    & 0.92 & 0.84      & 0.77   & 0.80 & 1.00        & 0.98     & 0.99  & 0.91 & 0.90 & 0.90  \\
k-NN                        & 0.91       & 0.94    & 0.92 & 0.85      & 0.79   & 0.82 & 1.00        & 1.00     & 1.00  & 0.92 & 0.91 & 0.91   \\
Naive Bayes                 & 0.79     & 0.98    & 0.88 & 0.91    & 0.44   & 0.59 & 1.00      & 0.98     & 0.99  & 0.90 & 0.80 & 0.82   \\
MLP                         & 0.81       & 0.92    & 0.86 & 0.75      & 0.54   & 0.63 & 0.98        & 0.98     & 0.98  &0.85 &0.81 &0.82    \\
SVM                         & 0.85       & 0.90    & 0.88 & 0.78      & 0.65   & 0.71 & 0.95        & 1.00     & 0.97  &0.86 &0.85 & 0.85    \\ 
\bottomrule
\end{tabular}
}
\end{table*}

The training set for Baltimore consisted of a random sample of 60\% of the data. The remaining 40\% was used to validate our prediction model. The vacant lots in Baltimore have three available conversions. QCMOS represented lots that were converted to qualified community open spaces, COMMUNITY GARDEN represented lots that were converted to community gardens, URBAN FARM represented lots that were converted to urban farms. The results are given in Table~\ref{tab:balti_threeclass}.

All the classifiers have significantly high precision and recall when it comes to classifying Urban Farms, while a lower precision and recall for trying to classify QCMOS. The primary reason for this may be because community managed open spaces are more ad hoc, as they are created by community members. In contrast, urban farms are only developed by the city and urban planners, and hence follows a stricter pattern.

\subsection{Vacant Lot Selection Between Two Cities}

For our cross city prediction model, we converted the dataset of Baltimore to point towards a binary class, indicating whether a vacant lot has been converted or not. The Philadelphia dataset already consists of binary classes, and as a result we performed two experiments. In the first one, we trained our model using data from Baltimore and performed predictions on Philadelphia. In the second one, we trained our model using data from Philadelphia and performed predictions on Baltimore. 

\begin{table}[!htb]
\centering
\caption{Prediction statistics with Baltimore as training set}
\label{tab:4.8}
\begin{tabular}{@{}llll@{}}
\toprule
\multirow{2}{*}{Classifier} & \multicolumn{3}{l}{ADOPTED} \\ \cmidrule(l){2-4} 
                            & Precision  & Recall  & F1   \\ 
\midrule
Random Forest               & 0.58       & 0.19    & 0.28 \\
k-NN                        & 0.55       & 0.40    & 0.46 \\ \bottomrule
\end{tabular}
\end{table}

\begin{table}[!htb]
\centering
\caption{Prediction statistics with Philadelphia as training set}
\label{tab:4.9}
\begin{tabular}{@{}llll@{}}
\toprule
\multirow{2}{*}{Classifier} & \multicolumn{3}{l}{ADOPTED} \\ \cmidrule(l){2-4} 
                            & Precision  & Recall  & F1   \\ 
\midrule
Random Forest               & 0.44       & 0.53    & 0.48 \\
k-NN                        & 0.46       & 0.51    & 0.48 \\ \bottomrule
\end{tabular}
\end{table}

The results of the experiments are given in Table~\ref{tab:4.8} and Table~\ref{tab:4.9}. As can be seen, the precision, recall and accuracy are significantly low for cross-city predictions i.e. they are no better than random. One possible reason could be that each city has their own vacant lot programs and as a result patterns do not necessarily match up. Another possible reason could be that the predictions made by our model could indicate vacant lots that have the indications of being a beneficial conversion, but simply has not been converted yet in another city. Further evaluation by expert urban planners and decision makers for our predictions would give a better indication.


\begin{table}[!htb]
\centering
\caption{Results of Cross City Experiments for {\rec}}
\label{tab:cross_city}
\begin{tabular}{@{}llll@{}}
\toprule
\multirow{2}{*}{Training Set}                                                 & \multicolumn{3}{l}{Adopt} \\ \cmidrule(l){2-4} 
                                                                              & Precision & Recall & F1   \\ \midrule
\begin{tabular}[c]{@{}l@{}}Baltimore: 100\%\\ Philadelphia: 25\%\end{tabular} & 0.55      & 0.55   & 0.55 \\ \hline
\begin{tabular}[c]{@{}l@{}}Baltimore: 100\%\\ Philadelphia: 50\%\end{tabular} & 0.57      & 0.54   & 0.56 \\ \hline
\begin{tabular}[c]{@{}l@{}}Baltimore: 50\%\\ Philadelphia: 100\%\end{tabular} & 0.81      & 0.90   & 0.85 \\ \hline
\begin{tabular}[c]{@{}l@{}}Baltimore: 25\%\\ Philadelphia: 100\%\end{tabular} & 0.75      & 0.84   & 0.79 \\ \bottomrule
\end{tabular}
\end{table}

We further experimented with mixed datasets, where we included certain percentage of the test city dataset in our training set to capture some of the characteristics of the test city and predict the vacant lot selections in it. As can be seen in Table~\ref{tab:cross_city}, utilizing 100\% of the Baltimore dataset and predicting on Philadelphia did not give an adequate result. However, if we utilized only 25\% of the Baltimore dataset along with 100\%of the Philadelphia dataset, the precision and recall is significantly higher than random selection. 

As can be seen in Figure~\ref{fig:balti_test}, the predicted and observed values for vacant lot status for our training does not differ much. In fact, most of our misses in that particular area is because {\rec} is recommending vacant lots to be converted that have not been converted yet. They may be converted in the future, and as mentioned in the previous section further evaluation with actual urban planners may resolve this issue.

In Section~\ref{sec:intro}, we outlined two research questions for this paper. Throughout the previously mentioned experiments we tested and evaluated different aspects of the research questions. For \textbf{RQ1}, we discovered that we can predict whether a vacant lot should be converted within a single city effectively using either 1-NN or a Random Forest classifier with 8 decision trees. For vacant lot conversions between cities, having a certain percentage of data from the target city improves {\rec}'s learning. For \textbf{RQ2}, we discovered that for Baltimore, {\rec} provided an accurate classification of what a vacant lot should be converted to with a high F1 score.

\section{Future Work}\label{sec:future}
The primary contribution of this work is to build and analyze a vacant lot model that can be used to predict future vacant lot conversions based on historical conversions of vacant lots. Our prediction models can be optimized further and integrated as a part of a larger urban planning system. The goal of our recommendation system is not to provide a complete solution but to be a part of a larger tool that would help support decision making for cities. Furthermore, our model can also be deployed as a part of a vacant lot toolkit, that would recommend to members of the community on vacant lots that may have a greater impact if they adopt or convert it.

Our research can be further extended with the use of a greater number of cities, as we limited the scope of our research to only two city datasets. Furthermore, there is scope for improvement in our prediction model with the use of other non-stationary determinants such as satellite imagery, geographical weights etc. Our system can also be further evaluated by validating predictions made by expert urban planners, who can assess which vacant lots will have the most impact. In the future, it would also be an interesting project to attempt to build a generalized prediction model, instead of models specific to cities. 

\section{Conclusion}\label{sec:conclusion}
The objectives of this research were the design and development of a vacant lot model, build a dataset utilizing the proposed model, create {\rec} a recommender for urban vacant lot that utilizes the proposed model and dataset. The vacant lot model we built consisted of determinants such as distance to nearest public infrastructure, crime density, access through public transit, zoning policies etc.

We built our model for two example datasets, for the cities of Baltimore and Philadelphia, and built our prediction models for each of these cities on a portion of the datasets. We validated them against the remaining dataset, and found that our model captured vacant lot determinants and impact extensively well for each city, and performed relatively well across cities provided there were some training data from the target city. {\rec} also displayed that it can capture what a vacant lot should be converted to accurately as well. We also discovered that Random Forest with 8 decision trees and 1-NN classifiers performed significantly better than other classifiers. 

{\rec} displayed that it's feasible to have automatic recommendations as a starting point for tackling the vacant lot problem. Community leaders can use {\rec} to pick a vacant lot to convert, while urban planners can use our system for a more macro level approach in terms of targeting specific vacant lots. 







\appendix

\bibliographystyle{ACM-Reference-Format}
\bibliography{main} 

\end{document}